%% file: lctoppi0.tex
\documentclass[twocolumn,showpacs,superscriptaddress,amsmath,amssymb]{revtex4-1}
\usepackage[colorlinks,linkcolor=blue,anchorcolor=blue,citecolor=blue,urlcolor=blue]{hyperref}
\usepackage{subfigure}
\usepackage{graphicx}
\usepackage{array}
\usepackage{xcolor}
\usepackage{multirow}
\usepackage{overpic}

\usepackage{lineno}
\usepackage{LamcPPi0-defs}

\begin{document}
\normalsize
\parskip=5pt plus 1pt minus 1pt

\lefthyphenmin=5
\righthyphenmin=5

\title{Evidence of the Singly Cabibbo Suppressed decay $\ensuremath{\Lambda_c^+\to p\pi^0}$}

\input{authorlist_2023-05-06}

\date{\today}

\begin{abstract}

Evidence for the singly Cabibbo suppressed decay $\ensuremath{\Lambda_c^+\to p\pi^0}$ is reported for the first time with a statistical significance of $3.7\sigma$ based on 6.0~$\ensuremath{\mathrm{fb}^{-1}}$ of $\ensuremath{e^+e^-}$ collision data collected at  center-of-mass~energies between 4.600 and 4.843~GeV with the BESIII detector at the BEPCII collider.
The absolute branching fraction of $\ensuremath{\Lambda_c^+\to p\pi^0}$ is measured to be $(1.56^{+0.72}_{-0.58}\pm0.20)\times 10^{-4}$. 
Combining with the branching fraction of $\ensuremath{\Lambda_c^+\to n\pi^+}$, $(6.6\pm1.3)\times10^{-4}$, the ratio of the branching fractions of $\ensuremath{\Lambda_c^+\to n\pi^+}$ and $\ensuremath{\Lambda_c^+\to p\pi^0}$ is calculated to be $3.2^{+2.2}_{-1.2}$. As an important input for the theoretical models describing the decay mechanisms of charmed baryons, our result indicates that the non-factorizable contributions play an essential role and their interference with the factorizable contributions should not be significant. 
In addition, the absolute branching fraction of $\ensuremath{\Lambda_c^+\to p\eta}$ is measured to be $(1.63\pm0.31_{\rm stat}\pm0.11_{\rm syst}) \times10^{-3}$. 

\end{abstract}

\maketitle

The hadronic weak decays of the $\Lamcp$ baryon have attracted strong theoretical interest.  
However, till now there is no reliable phenomenological model which could clarify the complicated underlying dynamics of charmed baryon decays. 
The hadronic decay amplitudes of charmed baryons generally consist of factorizable and non-factorizable contributions.
Unlike charmed mesons, the non-factorizable contribution is not negligible compared to the factorizable ones since the $W$-exchange process is no longer subject to helicity and color suppression~\cite{Cheng2018,Cheng2015}. 
This makes the corresponding theoretical calculation more complicated. 
Moreover, the interference between the $W$-emission and exchange contributions 
remain unclear despite of the extensive studies by various phenomenological models.
Therefore, experimentally investigating the non-leptonic weak hadronic decays of $\Lamcp$ are highly desired, in which the two-body singly Cabibbo suppressed (SCS) decays, $\LamCNPi$, $\Lamcppi$, and $\Lamcpeta$, are the channels of most interest. Specifically, $\LamCNPi$ and $\Lamcppi$ receive the external and color-suppressed internal $W$-emission contributions, respectively, 
together with the non-factorizable $W$-exchange ones. 
In addition, the ratio of their branching fractions (BFs) is expected to be less sensitive to the input parameters in the phenomenological models, due to the significant cancellation of the corresponding correlated uncertainties, and is therefore effective to test the different models.

The measurements of $\Lamcpeta$~\cite{Ablikim2017,Li2021a} have been demonstrated to be consistent with theoretical calculations~\cite{Geng2019,Cheng2018} and the knowledge of non-factorizable contributions has improved. 
However, a long-standing contradiction between experimental results and predictions on $\Lamcppi$ exists.
Before 2019, the SU(3) flavor (SU(3)$_{f}$) symmetry models neglected the irreducible representation $\mathcal{O}(\overline{15})$
and the predicted BFs were consistent with measurements of Cabibbo favored processes. However, the predictions on
the BF of $\Lamcppi$ ~\cite{Geng2018,Lue2016}
are significantly greater than the upper limits measured by BESIII~\cite{Ablikim2017} and Belle~\cite{Li2021a} experiments.
Having taken $\mathcal{O}(\overline{15})$ contribution into account, and undergone some improvements, SU(3)$_{f}$ predictions updated the results for \Lamcp{} decays in 2019~\cite{Geng2019}.
Particularly, 
it successfully predicted the BF of \LamCNPi{}, which is measured by BESIII in 2022~\cite{Ablikim2022}.
However, the updated prediction on BF of  $\Lamcppi$, ($1.3\pm0.7) \times 10^{-4}$, 
still appears to be greater than the upper limit measured by Belle, $<8.0\times10^{-5}$, despite the large uncertainty. 
Additionally, by combining the results from Belle~\cite{Li2021a} and BESIII~\cite{Ablikim2022}, the ratio of the BFs between $\LamCNPi$ and $\Lamcppi$ is calculated to be greater than 7.2 at the 90\% confidence level.
This result unexpectedly conflicts with most theoretical predictions,
such as 2.0 
with the SU(3)$_{f}$ symmetry~\cite{Geng2018,Sharma1997,Lue2016}, 
4.7 with the SU(3)$_{f}$ symmetry including the contribution from $\mathcal{O}(6)$ and $\mathcal{O}(\overline{15})$~\cite{Geng2019},
4.5 
from the constituent quark model~\cite{Uppal1994}, and 3.5 from the dynamical calculation based on the
pole model and current algebra~\cite{Cheng2018}.
Very recently, Ref.~\cite{Zhong2023} calculated the BF of $\Lamcppi$ to be $(0.51_{-0.61}^{+0.59})\times10^{-4}$ 
or $(0.16\pm0.09)\times10^{-4}$,
 by considering SU(3) broken or SU(3) respected effect, respectively, 
 which is consistent with Belle's result
 and lower than previous SU(3)$_{f}$ predictions.
To clarify the aforementioned contradictions, searching for $\Lamcppi$ signal is crucial.
Moreover, the study of $\Lamcppi$ will definitely contribute to the interpretation of the non-factorizable contributions and their interferences with the factorizable components.

In this Letter, the first evidence of the SCS decay $\Lamcppi$ is reported using \ee{} collision data, collected by the BESIII detector at ten center-of-mass (c.m.)~energies between 4.600 and 4.843~GeV, corresponding to a total integrated luminosity of 6.0~fb$^{-1}$~\cite{Ablikim2022lum4,Ablikim2022lum4600}. 
These large data samples collected just above the $\Lamcp\Lamcm$ production threshold provide a clean environment and an excellent opportunity to search for $\Lamcppi$ with the double-tag approach. 
In addition, the decay $\Lamcpeta$ is also measured with the same approach to provide a validation. 
Throughout this Letter, charge-conjugate modes are implicitly included. 

The design and performance of the BESIII detector are described in detail in Ref.~\cite{Ablikim:2009aa}.
Simulated event samples are produced with a {\sc geant4}-based~\cite{Agostinelli:2002hh} Monte Carlo (MC) package, which includes the geometric description~\cite{Huang2022} of the BESIII detector and the detector response.
Signal MC samples of $\ee\to\Lamcp\Lamcm$, with \Lamcm{} decaying to nine specific tag modes (as described below and listed in Table~\ref{tab:yield-st-468}), 
together with $\Lamcp$ decaying to $p\piz$ or $p\eta$, are used to determine the detection efficiencies.  They are generated by {\sc kkmc} including the effects of initial-state radiation (ISR) and the beam energy spread. 
To estimate backgrounds, inclusive MC samples, which consist of \LCLC{} events, $D_{(s)}^{(*)}$ production, ISR production of vector charmonium(-like) states, Bhabha scattering, $\mu^+\mu^-$, $\tau^+\tau^-$, $\gamma\gamma$ events and other inclusive hadronic processes incorporated in {\sc kkmc}~\cite{Jadach:2000ir}, are generated.
Subsequent decays of all intermediate states are modeled with {\sc evtgen}~\cite{Lange:2001uf, Ping:2008zz} using the BFs either taken from the Particle Data Group~\cite{Workman2022}, when available, or modeled with {\sc lundcharm}~\cite{Chen:2000tv,Yang2014}. 
Final state radiation from charged final state particles is incorporated using the {\sc photos} package~\cite{Richter-Was:1992hxq}.

This work is carried out by the double-tag (DT) method. 
First, we select a data sample of the process $e^+e^- \rightarrow \Lamcp\Lamcm$,  called the single-tag (ST) sample,  by tagging a \Lamcm{} baryon with one of the nine exclusive hadronic decay modes, as listed in the first column of Table~\ref{tab:yield-st-468}.
Then, we search for the signal decays \Lamcppi{} and \Lamcpeta{} in the system recoiling against the ST \Lamcm{} candidates, referred to as the DT sample hereafter.

Charged tracks, photon candidates as well as the intermediate $\piz$, $\Ks$, $\bar{\Lambda}$, and $\bar{\Sigma}^{0}$ states are selected and reconstructed using the same criteria described in detail in Ref.~\cite{Ablikim2022}. 
In addition, the kaon candidate is now required to have a particle identification (PID) probability to be a kaon greater than 0.0005, in order to suppress background with pions misidentified as kaons.
PID for charged tracks combines measurements of the energy deposited in
the MDC, $dE/dx$, and the flight time from the TOF to form likelihoods $L(h)$ 
($h = p, K, \pi$) for each hadron hypothesis. 
Two variables, the energy difference, $\dE = E_{\Lamcm} - \Ebeam$, and the beam energy constrained mass $M_\mathrm{BC} = \sqrt{\Ebeam^2/c^4 - | \pALC |^2/c^2}$, are adopted to identify the ST \Lamcm{} candidates, where \Ebeam{} is the beam energy, $E_{\Lamcm}$ and \pALC{} are the energy and momentum of the \Lamcm{} candidate, respectively.
The tagged candidates are selected with the minimum $\left|\Delta E\right|$ among all the candidates, and are required to satisfy the $\dE$ requirements listed in the second column of Table~\ref{tab:yield-st-468}.
To avoid cross feed from the other ST modes, the same requirements as in Ref.~\cite{Ablikim2018} are applied for the $\Lamcm \to \apks\pi^0$, $\Lamcm\to\apks\pi^+\pi^-$ and $\bar{\Lambda}\pim\pip\pim$ modes.

The ST yields are extracted by performing unbinned maximum likelihood fits to the corresponding $M_\mathrm{BC}$ distributions in the range (2.2, $\Ebeam$)~$\gevcc$. 
In the fit, the signal shape is modeled by the \mBC{} spectrum extracted from the signal MC sample convolved with a Gaussian function to compensate the resolution difference between the data and the MC simulation.
The background is described by an ARGUS function~\cite{Albrecht1990} with the endpoint parameter fixed to the corresponding $\Ebeam$.
The detection efficiencies of ST candidates are estimated with MC samples of $\ee \to \Lamcp\Lamcm$ with $\Lamcm$ decaying to one of the nine tag modes and $\Lamcp$ decaying generically to all possible final states.
The ST yields and efficiencies at $\sqrt{s}=4.682$~GeV are shown in Table~\ref{tab:yield-st-468}. The results, including the ST yields and efficiencies for the other nine c.m.~energies, and the fits for all c.m.~energies are summarized in Supplemental Material~\cite{BESIII:Supplemental}. 
The total ST yield for all the ten c.m.~energies is $119398\pm413_{\rm stat}$ events.

\begin{table}[!htbp]
  \begin{center}
  \caption{The \dE{} requirements, the ST yields in data, and the detection efficiencies of the ST and DT candidates for nine tag modes at  $\sqrt{s}=4.682$~GeV. The uncertainties of the ST yields are statistical. }
  \renewcommand\arraystretch{1.2}
    \begin{tabular}{ l p{0.7cm}<{\raggedleft} @{,\, } p{0.3cm}<{\raggedright} p{0.8cm}<{\raggedleft} @{ $\pm$ } p{0.5cm}<{\raggedright} p{0.9cm}<{\centering} p{1.2cm}<{\centering} p{1cm}<{\centering} }
      \hline
      \hline
	     & \multicolumn{2}{c}{\multirow{2}{*}{$\dE$ (MeV)}} & \multicolumn{2}{c}{\multirow{2}{*}{$N_{i}^{\mathrm{ST}}$}} & $\epsilon_{i}^{\mathrm{ST}}$ & $\epsilon_{i}^{\mathrm{DT}}$ & $\epsilon_{i}^{\mathrm{DT}}$ \\
	      &\multicolumn{2}{c}{}  & \multicolumn{2}{c}{}   & (\%) &  $p\piz$ (\%) &   $p\eta$ (\%)\\
      \hline
	                    
	    $\apkpi$                       & $(-34$ & $20)$  &17557 & 149      &   47.03 & 22.13& 20.64 \\
            $\apks$                        & $(-20$ & $20)$  &3486  & 62       &   49.61 & 24.66& 22.43 \\
            $\apkpi\pi^0$                  & $(-30$ & $20)$  &2087  & 48       &   41.55 & 21.00& 19.23 \\
            $\apks\pi^0$                   & $(-30$ & $20)$  &4159  & 100      &   14.97 & 7.83 &  7.12 \\
            $\apks\pi^+\pi^-$              & $(-20$ & $20)$  &1545  & 53       &   17.28 & 7.73 &  8.17 \\
            $\bar{\Lambda}\pi^-$           & $(-20$ & $20)$  &3776  & 75       &   15.36 & 7.83 &  7.53 \\
            $\bar{\Lambda}\pi^-\pi^0$      & $(-30$ & $20)$  &1352  & 50       &   18.21 & 7.45 &  7.72 \\
            $\bar{\Lambda}\pi^-\pi^+\pi^-$ & $(-20$ & $20)$  &1858  & 52       &   12.66 & 5.14 &  5.64 \\
            $\bar{\Sigma}^0\pi^-$          & $(-20$ & $20)$  &1084  & 36       &   20.09 & 11.48& 10.43 \\
      \hline\hline
    \end{tabular}
          \label{tab:yield-st-468}
  \end{center}
\end{table}

The signal decays $\Lamcppi$ and $\Lamcpeta$ are searched for in the remaining objects recoiling against the ST $\Lamcm$ candidates.
The selection criteria for charged tracks and photons are the same as those in the ST selection. 
The distance of the closest approach to the interaction point along the beam direction is required to be less than $10$ cm and $20$ cm for tight and loose tracks, respectively~\cite{Ablikim2022}. 
In addition, the corresponding distance perpendicular to the beam direction is required to be less than $1$ cm for a tight track.
To suppress multi-track backgrounds, there must be only one tight track, identified as a proton, and no loose tracks. 
In order to eliminate noise created by $\bar{p}$ in the electromagnetic calorimeter (EMC), photons are required to be separated from $\bar{p}$ with an opening angle greater than  $30^{\circ}$.
To improve its purity, the photon candidates are further required to have  their lateral moment in the range of $(0.05, 0.40)$ and have an $E_{3\times3}/E_{5\times5}$ greater than 0.85, where $E_{3\times3}$ ($E_{5\times5}$) is the shower energy summed over $3\times3$ ($5\times5$) group of crystals around the central seed crystal.  Events with at least two photon candidates are kept for further analysis.

The signal candidate is reconstructed with the selected proton and two photons. For those events with multiple candidates (roughly 10\% of the total events), only the one with the minimum energy difference  $|\dE_{p2\gamma}| = |E_p+E_{\gamma 1}+E_{\gamma 2}-\Ebeam|$ is kept, where $E_p$ and $E_{\gamma 1/2}$ are the energies of the proton and two photons, respectively.
The criterion $-0.080<\dE_{p2\gamma}<0.035$~GeV, optimized by using the inclusive and signal MC samples, is applied to further suppress the background.

A clear $\Lambda$ peak is observed in the invariant mass distribution of the proton from the signal side and a $\pim$ from the ST side.  
Thus, an event is rejected if the invariant mass of any such combination of $p$ and $\pim$ lies within (1.111, 1.121)~$\gevcc$.
Similarly, an $\omega$ signal is seen in the invariant mass distribution of the $\piz$ from the signal side and $\pip\pim$ pairs from ST side. Events with an invariant mass of any such combination within (0.733, 0.833)~$\gevcc$ are also rejected.  

\begin{figure}[htbp]
\hspace{-5mm}
\subfigure{
  \begin{overpic}
    [width=0.248\textwidth, trim=5 0 0 0, clip]{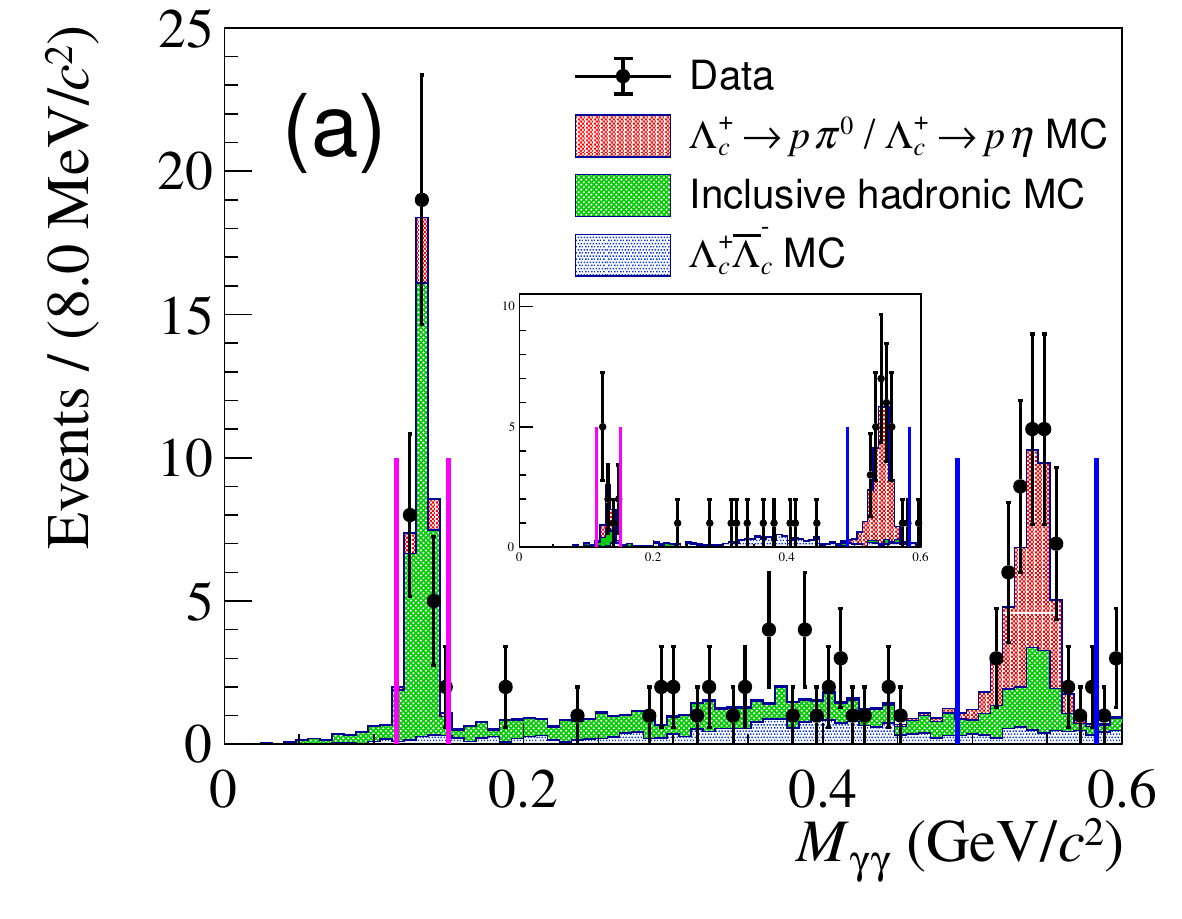}
  \end{overpic}
    \label{fig:myy}
}
\hspace{-5mm}
\subfigure{
  \begin{overpic}
    [width=0.248\textwidth, trim=5 0 0 0, clip]{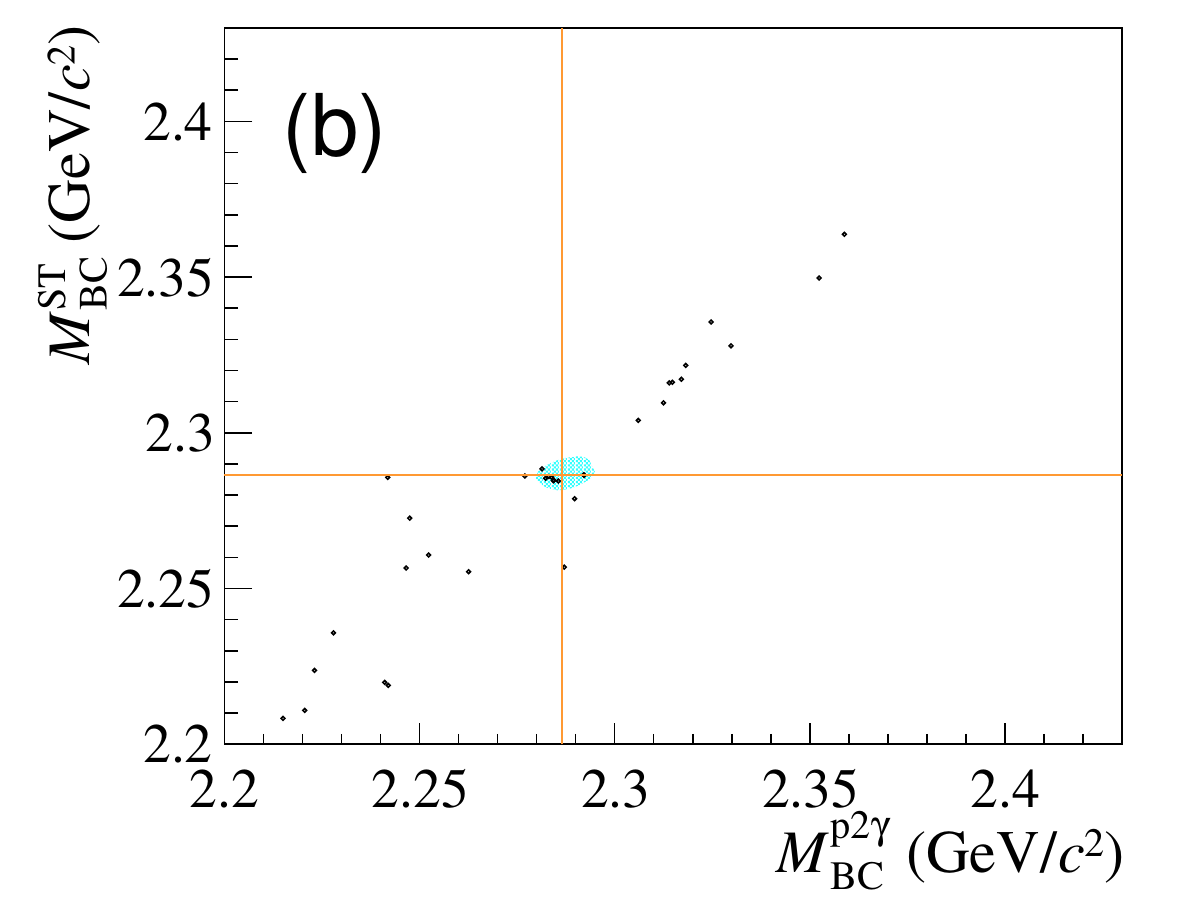}
  \end{overpic}
    \label{fig:2Dpi0}
}

\hspace{-5mm}
\subfigure{
  \begin{overpic}
    [width=0.248\textwidth, trim=5 0 0 0, clip]{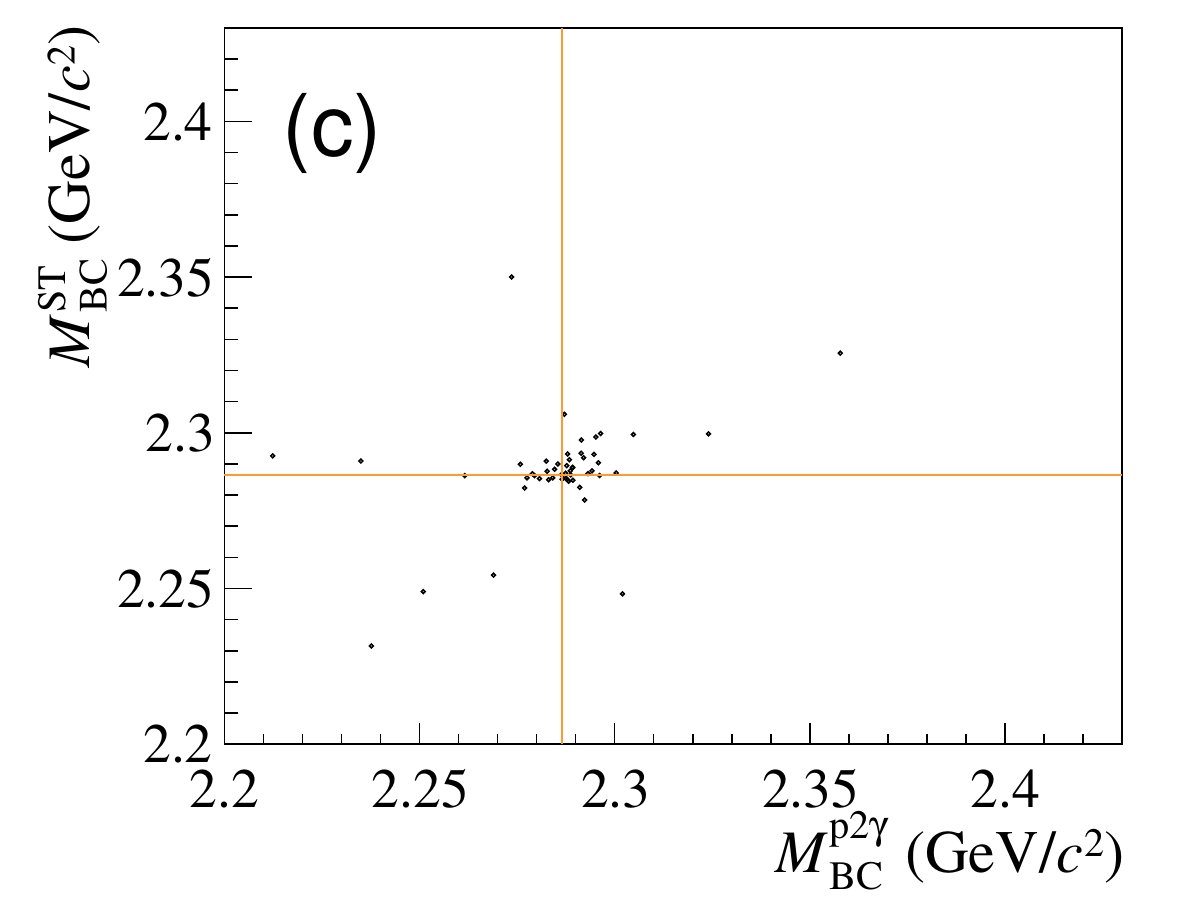}
  \end{overpic}
    \label{fig:2Deta}
}
\hspace{-5mm}
\subfigure{
  \begin{overpic}
    [width=0.248\textwidth, trim=5 0 0 0, clip]{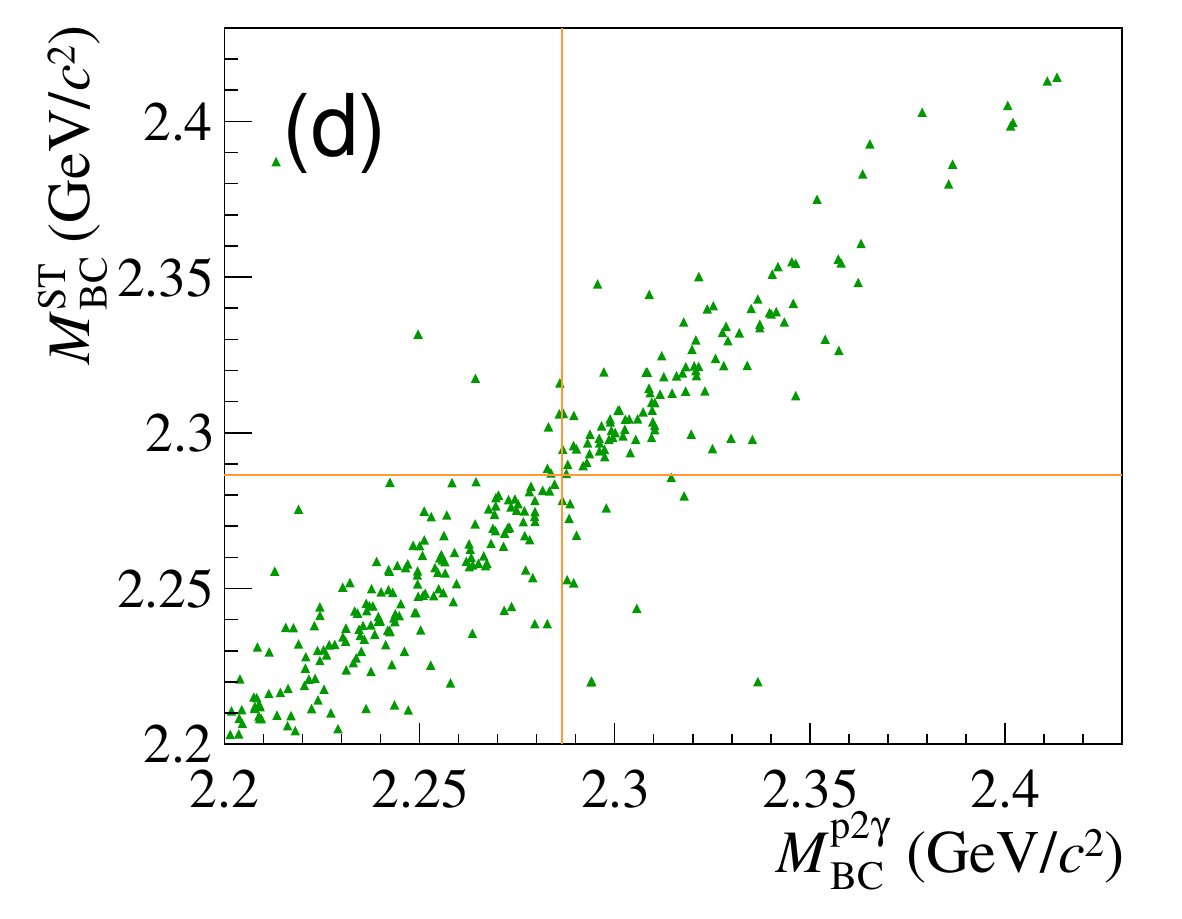}
  \end{overpic}
    \label{fig:2Dsideband}
}

    \caption{ The invariant mass distribution of $\gamma\gamma$ (a). In the subfigure shows the $M_{\gamma\gamma}$ distribution in 2D \mBC{} signal region ($M_{\mathrm{BC}}^{\mathrm{ST}} \in [2.280,2.295]$ GeV/$c^{2}$ and $M_{\mathrm{BC}}^{p2\gamma} \in [2.280,2.295]$ GeV/$c^{2}$).
The scatter plots of $\dtMBC$ versus $\stMBC$ for the candidate events with $M_{\gamma\gamma}$ within  (b) $\piz$ signal region of data, (c) $\eta$ signal region of data, and (d) $\piz$ signal region of inclusive hadronic background MC sample whose luminosity is 10 times of data. The blue oval in (b) represents the one-sigma resolution of signal MC. The orange lines denote the nominal $\Lambda_c$ mass.}
\label{fig:distribution}
\end{figure}

The signal decays are examined by \stMBC{} of the tag side, \myy{} and \dtMBC{} of the signal side.
Here, \myy{} is the invariant mass of two photons, and \stMBC{} and \dtMBC{} are the beam energy constrained masses of the ST $\Lamcm$ and signal candidate.
The combined distribution of \myy{} from the ten c.m.~energies is shown in Fig.~\ref{fig:myy}, where both $\piz$ and $\eta$ signals are observed. 
The $\piz$ and $\eta$ signal regions are defined as $(0.115,0.150)$~$\gevcc$ and $(0.490,0.583)$~$\gevcc$, as indicated by the regions between two magenta and blue lines, respectively. 
The background in region $(0.17,0.47)$~$\gevcc$ is dominated by the decays $\Lamcp\to \Sigma^+(\to p\piz)\piz$ and $\Lamcp\to p\Ks (\to \piz\piz)$.
The $\eta$ signal is dominated by $\Lamcp\to p\eta$; however, the $\piz$ signal suffers from a large contamination from inclusive hadronic processes. 

\begin{figure*}[htbp]
\subfigure{
  \begin{overpic}
    [width=0.37\textwidth, trim=5 0 0 0, clip]{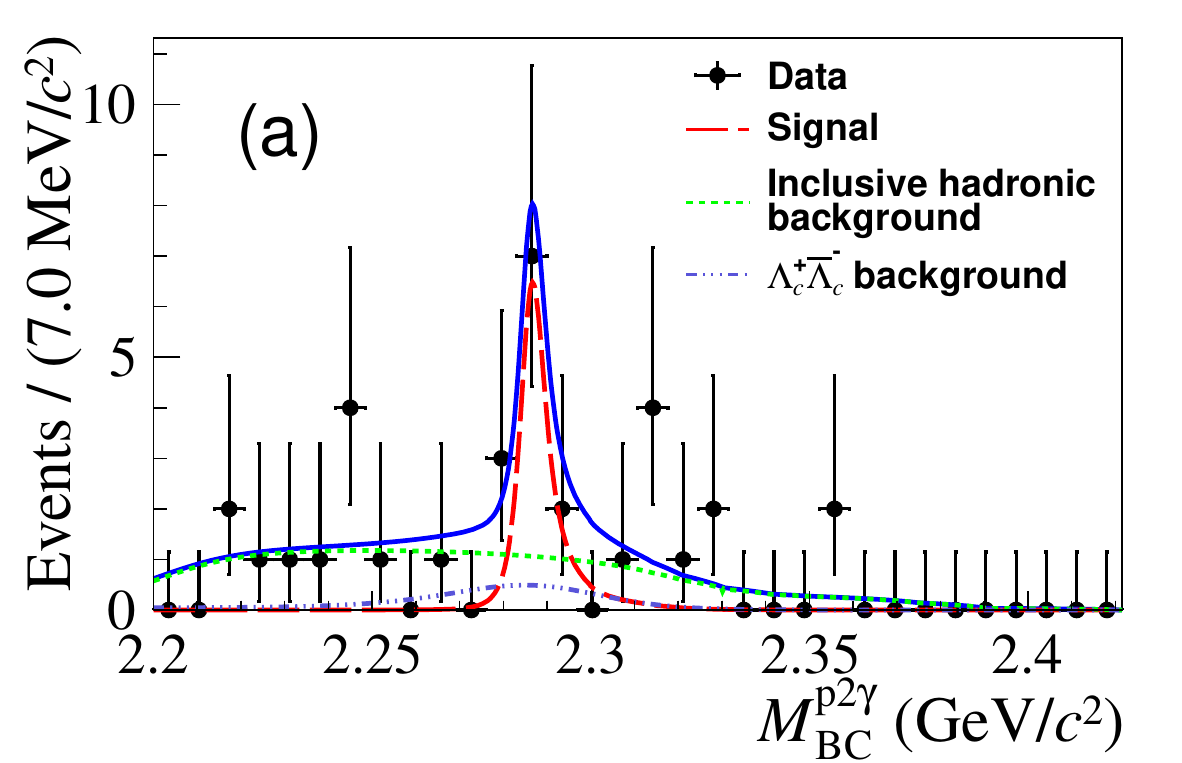}
  \end{overpic}
    \label{fig:pi0sig}
}
\subfigure{
  \begin{overpic}
    [width=0.37\textwidth, trim=5 0 0 0, clip]{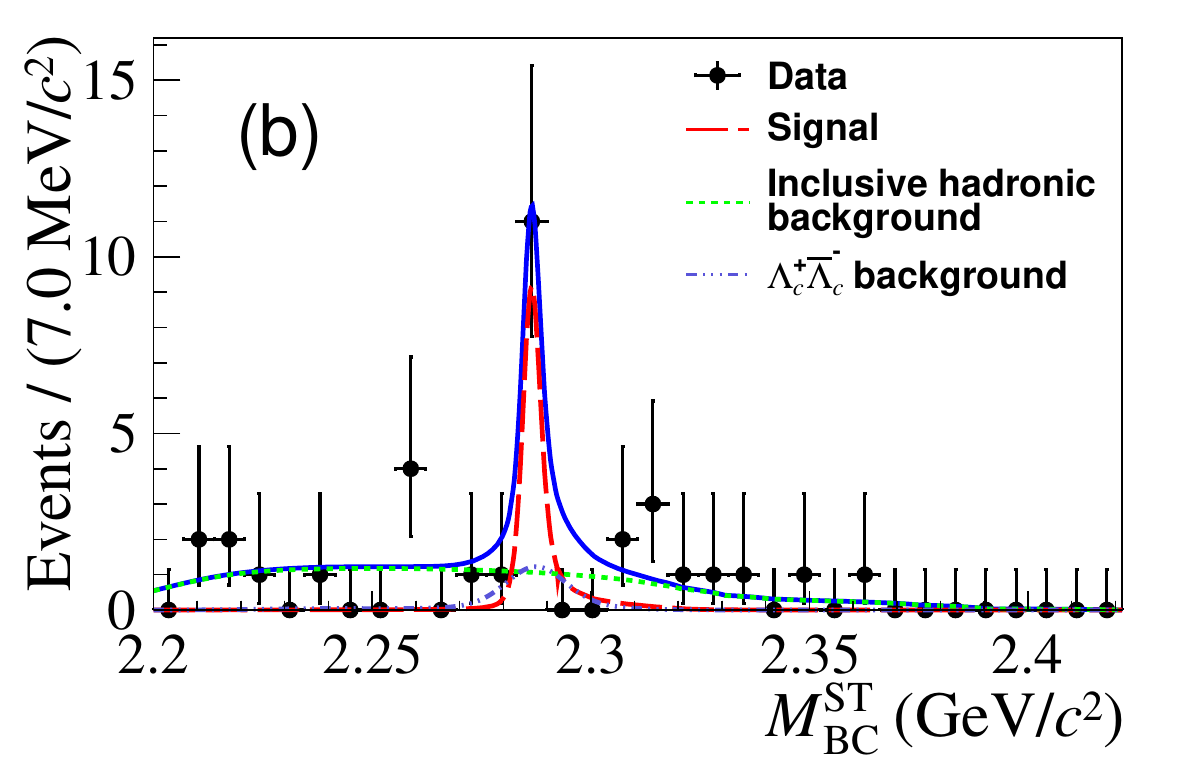}
    \label{fig:pi0tag}
  \end{overpic}
}
\subfigure{
  \begin{overpic}
    [width=0.37\textwidth, trim=5 0 0 0, clip]{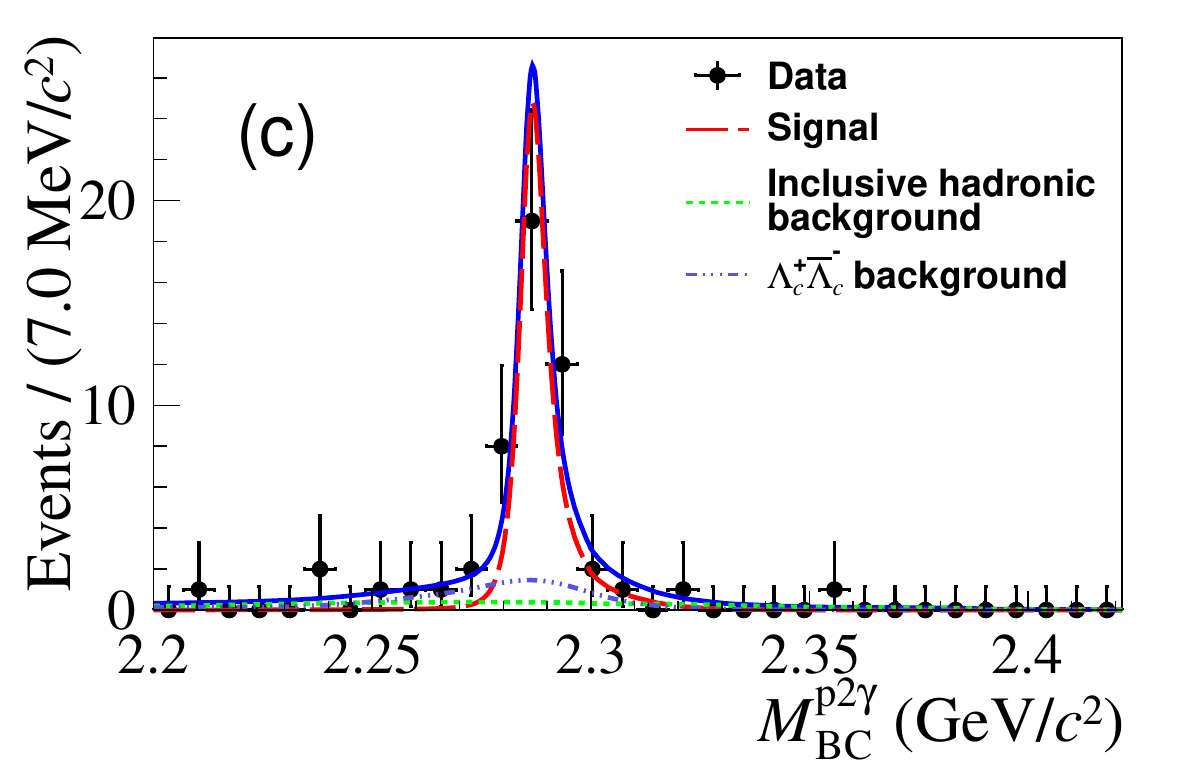}
  \end{overpic}
    \label{fig:etasig}
}
\subfigure{
  \begin{overpic}
    [width=0.37\textwidth, trim=5 0 0 0, clip]{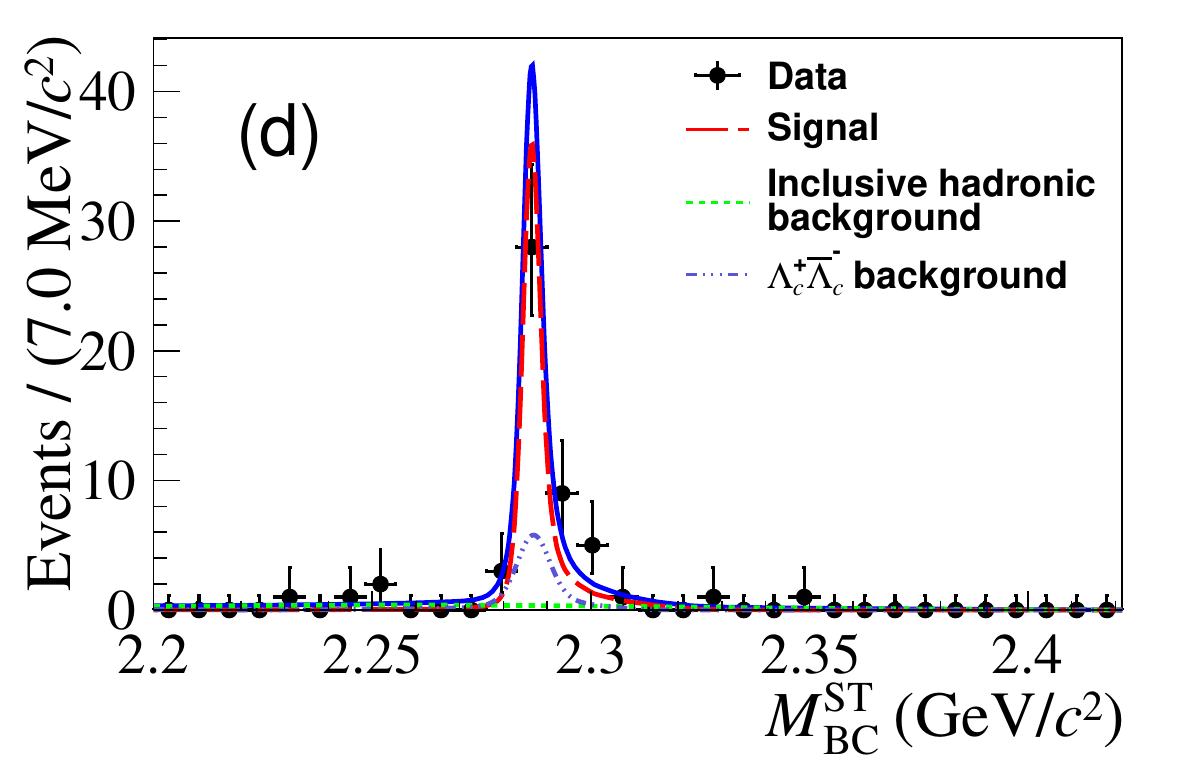}
  \end{overpic}
    \label{fig:etatag}
}

\caption{The 1D projections of the 2D fits for \Lamcppi{} and \Lamcpeta{}, where (a) and (b) are $\dtMBC$ and $\stMBC$ for $\Lamcppi$, and (c) and (d) are the corresponding plots for $\Lamcpeta$. 
The black dots denote data and the blue solid curves are the sum of the fit functions. The signal is illustrated by the red dashed curves, while background from the inclusive hadronic events and the non-signal $\ee \to \Lamcp\Lamcm$ events are denoted by the green and blue dashed curves, respectively.}
\label{fig:fitpi0}
\end{figure*}

The distribution of \stMBC{} versus \dtMBC{} and their projections for the candidate events with \myy{} in the $\piz$ signal region, are shown in Fig.~\ref{fig:2Dpi0} and Fig.~\ref{fig:pi0sig}, \ref{fig:pi0tag}, respectively. 
The events accumulating around the intersection of  $\stMBC=M_{\Lambda_c}$ and $\dtMBC=M_{\Lambda_c}$ provide evidence for $\Lamcppi$ (here, $M_{\Lambda_c}$ denotes the known $\Lambda_c$ mass~\cite{Workman2022}). 
Similarly, Fig.~\ref{fig:2Deta} and Fig.~\ref{fig:etasig}, \ref{fig:etatag} illustrate the distribution of \stMBC{} versus \dtMBC{} and their projections with \myy{} in the $\eta$ signal region, respectively, where the $\Lamcpeta$ signal can be clearly seen.
The distribution of inclusive hadronic background MC sample in $\piz$ signal region is shown in Fig.~\ref{fig:2Dsideband}, where no accumulation of events in the vicinity of the intersection point is observed.
As is shown in Fig.~\ref{fig:distribution}, 9 events are observed in one-sigma signal region,  
while inclusive hadronic MC sample predicts about 0.7 event for the data equivalent luminosity. 
The distribution of the $\Lamcp\Lamcm$ background MC sample in the $\piz$ signal region as well as those in the $\eta$ signal region are given in the Supplemental Material~\cite{BESIII:Supplemental}. 

The BFs of the decays $\Lamcpeta$ and $\Lamcppi$   are obtained by performing a simultaneous unbinned maximum-likelihood fit on the 2-dimensional (2D) distribution of $\dtMBC$ versus $\stMBC$ among the ten data samples with different c.m.~energies.
In the fit, the signal is modeled with an MC-simulated shape convolved with  Gaussian functions (for  $\dtMBC$ and $\stMBC$, individually) representing the resolution difference between data and MC simulation.
The mean and width of the Gaussian functions are extracted from the fit to a control sample of $\Lamcp\to \pkpi\pi^0$.
The inclusive hadronic background is modeled with the product of two ARGUS functions and a Student's $t$-distribution describing the dispersion of the inclusive hadronic background in the diagonal direction.
Details of the background functions as well as the free and fixed parameters are given in the Supplemental Material~\cite{BESIII:Supplemental}.
The background from non-signal $\ee \to \Lamcp\Lamcm$ events is modeled with the inclusive MC samples.
The same decay BF is shared among the different c.m.~energies in the simultaneous fit with the relation
 \begin{equation} \label{eq:br}
  \br \cdot {\Sigma_{i}\, (\effdouble \cdot \Nsingle / \effsingle) } + N_{\rm bkg} = N_{\mathrm{total}},
\end{equation}
where the subscript $i$ represents the $i$-th ST mode, $N_{\mathrm{total}}$ and $N_{\rm bkg}$ denote the total event yields and backgrounds,  
\effdouble{} represents the DT detection efficiencies, which are extracted from the DT signal MC samples.
The quantities \Nsingle{}, \effsingle{} and \effdouble{} used in the 2D fit are summarized in Table~\ref{tab:yield-st-468} for the data sample with c.m.~energy $\sqrt{s} = 4.682$~GeV, and in Supplemental Material~\cite{BESIII:Supplemental} for the other nine c.m.~energies. 

The simultaneous fit yields the BFs $\mathcal{B}(\Lamcpeta)=(1.63\pm0.31)\times 10^{-3}$ and $\mathcal{B}(\Lamcppi)=(1.56^{+0.72}_{-0.58})\times 10^{-4}$,  which correspond to the signal yields of $34.7\pm6.6$ and $8.8^{+4.0}_{-3.3}$ events, respectively, for the total ten c.m.~energies.
The projections of  $\dtMBC$ and $\stMBC$ are shown in  Fig.~\ref{fig:fitpi0}, where the fit curves describe the data well.
The statistical significances, which are estimated from the change of likelihood values and the change of degrees of freedom with and without the signal function included in the fit, are $6.9\sigma$ and  $3.8\sigma$ for  \Lamcpeta{} and  \Lamcppi{}, respectively.

The systematic uncertainties for the BF measurement include those associated with the ST and DT yields, the detection efficiencies from the signal side, the BFs of $\piz\to\gamma\gamma$ and $\eta\to\gamma\gamma$, and MC statistics.  

The uncertainties in the ST detection efficiencies are cancelled by the DT method.
The uncertainty associated with DT efficiencies, $\effdouble$, has several sources.  
Systematic effects due to the proton tracking and PID efficiencies are estimated using the control sample $\jpsi \to p\bar{p}\pip\pim$~\cite{Ablikim2022} and give 1.0\% and 1.0\%, respectively.
The uncertainties originating from the photon reconstruction efficiency and shower shape requirements are studied by the control sample  $\jpsi \to \rho^0 \piz \to \pip \pim\piz$~\cite{Ablikim2010a}, where 1.0\% and 0.5\% per photon are obtained, respectively.
The uncertainty from the requirement of the opening angle between photon and anti-proton is 0.5\% per photon, which is obtained by studying a control sample of $J/\psi \to p\bar{p}\piz$.
A Barlow Test~\cite{Barlow2002} is carried out to examine the uncertainties due to the mass window requirements vetoing the backgrounds associated with $\Lambda$ and $\omega$, where no significant systematic deviation is observed. 
The systematic uncertainty from the $\dE$ requirement on the signal side is found to be 0.3\% by studying the control samples of modes $\Lamcp \to \pkpi \piz$ and $\Lamcp \to \Lambda\pip\piz$.
The uncertainty due to the signal MC model is taken as 0.4\%~\cite{BAM580arxiv}. 
Uncertainties from the BFs of $\eta\to\gamma\gamma$ and $\piz\to\gamma\gamma$ are taken from PDG~\cite{Workman2022}. 
The uncertainty due to the MC statistics is 0.1\%. 
The ST yields contribute to the uncertainty of BF by 0.5\%~\cite{Ablikim2022}. 
The uncertainties due to the DT yields extraction are 5.7\% and 12.4\% for $\Lamcpeta$ and $\Lamcppi$, respectively, which is the quadratic sum of the individual changes of signal yields obtained from the alternative fits by changing the fixed parameters in the shapes of the signal and background by $\pm 1\sigma$.

Assuming all sources of the uncertainties are uncorrelated, the quadratic sums of the different contributions are taken as the total uncertainties.
For \Lamcpeta{} and \Lamcppi{}, the total systematic uncertainties become 6.4\% and 12.8\%, respectively.
Alternative fits of changing the fixed parameters of background shape with $\pm 1\sigma$ deviation are performed individually, and corresponding fits with signal removed are performed. 
The most conservative significance among all changes is taken as the statistical significance accounting for the systematic uncertainty, and the significance is 3.7$\sigma$.

\begin{figure}[!htp]
    \begin{center}
        \includegraphics[width=0.44\textwidth, trim=5 0 0 0, clip]{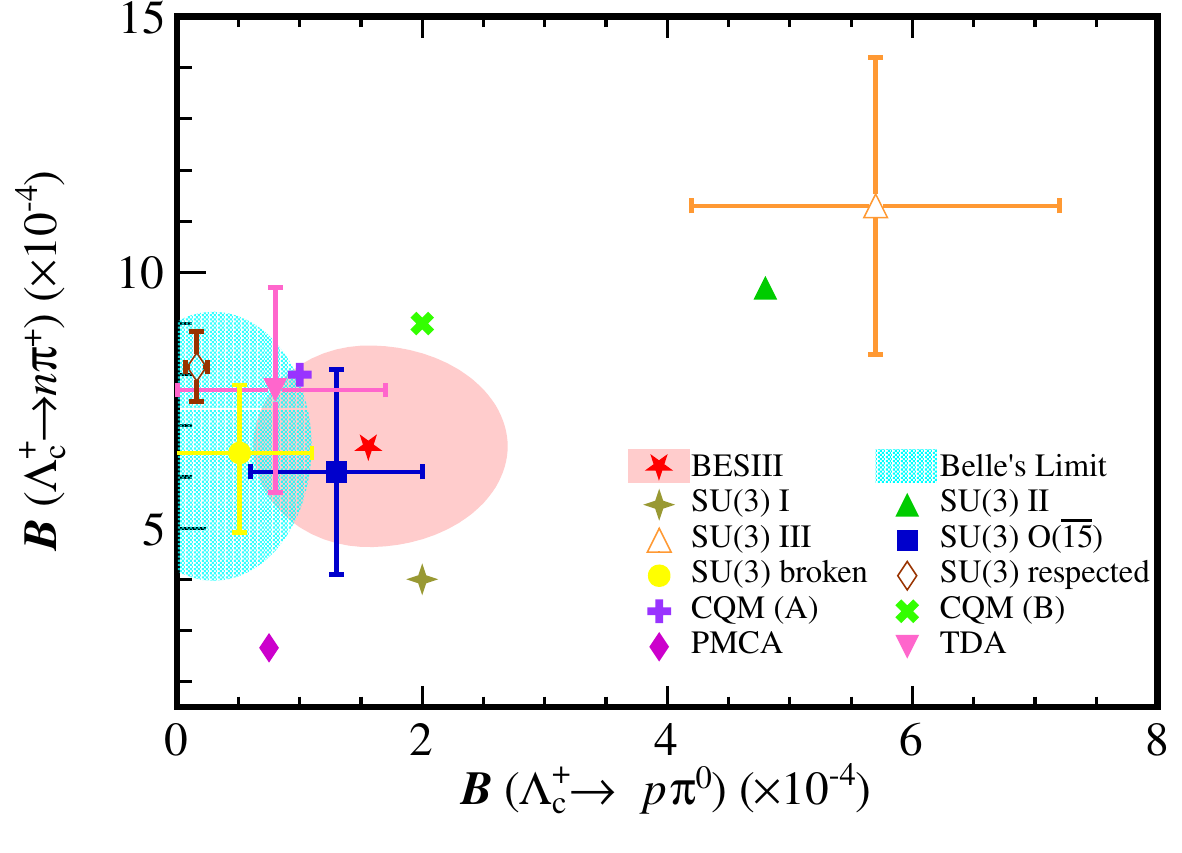}
    \end{center}
    \caption{
	    Distribution of BF of \LamCNPi{} versus BF of $\Lamcp\to p\pi^0$. 
	    The red star denotes the results measured by this work and the
	    pink contour corresponds to the 68\% confidence interval of the results. 
	    The cyan shadowed area shows the upper limit at 90\% confidence level from Belle~\cite{Li2021a}.
	    The BF of \LamCNPi{} of both two shadowed area is from BESIII~\cite{Ablikim2022}.
	    Other symbols stand for the phenomenological predictions:  SU(3)$_{f}$ symmetry 
(SU(3) I refers to Ref.~\cite{Sharma1997}, SU(3) II refers to Ref.~\cite{Lue2016}, and SU(3) III refers to Ref.~\cite{Geng2018}),
SU(3) including the contribution from $\mathcal{O}(6)$ and $\mathcal{O}(\overline{15})$ (referred to as SU(3) $\mathcal{O}(\overline{15})$)~\cite{Geng2019}, 
SU(3) broken and respected~\cite{Zhong2023}, 
constituent quark model (CQM)~\cite{Uppal1994} with two predictions (A) and (B), dynamical calculation based on the pole model and current algebra (PMCA)~\cite{Cheng2018}, and topological-diagram approach (TDA)~\cite{Zhao2020}. 
}
     \label{fig:BF_ratio_compare}
\end{figure}

In summary, based on \ee{} collision data samples with a total integrated luminosity of 6.0~\ifb{} collected at c.m.~energies between 4.600 and 4.843~GeV with the BESIII detector, the SCS decays \Lamcppi{} and \Lamcpeta{} are measured using a DT method.
Evidence is obtained for the decay \Lamcppi{}, with a statistical significance of $3.7\sigma$ and a decay BF of $(1.56^{+0.72}_{-0.58}\pm0.20)\times 10^{-4}$. 
The BF of \Lamcpeta{} is measured as $(1.63\pm0.31_{\rm stat}\pm0.11_{\rm syst})\times 10^{-3}$, consistent with previous results~\cite{Ablikim2017,BAM580arxiv,Li2021a}. 
The $\mathcal{B}(\Lamcppi)$ result distinctly exceeds the upper limit measured by Belle experiment.
Taking $\mathcal{B}(\LamCNPi) = (6.6\pm1.3)\times10^{-4}$~\cite{Ablikim2022}, the ratio of the BFs between \LamCNPi{} and $\Lamcp\to p\pi^0$ is calculated to be $3.2^{+2.2}_{-1.2}$, where $3.2$ denotes the most probable value, and the upper and lower errors covers the 68\% confidence interval. 
This value is obtained by generating toy MC samples and forming a Probability Density Function of the ratio. The detailed method is given in the Supplemental Material~\cite{BESIII:Supplemental}.
This ratio is consistent with majority of phenomenological predictions~\cite{Sharma1997,  Lue2016, Uppal1994, Cheng2018, Geng2018, Geng2019}. More importantly, not only the ratio but also the individual BFs of
$\Lamcppi$ and $\LamCNPi$ agree with the calculation of SU(3)$_{f}$ symmetry including the 
contributions from both  $\mathcal{O}(6)$ and $\mathcal{O}(\overline{15})$~\cite{Geng2019}, 
which indicates the non-factorizable contributions play an essential role in these two decays 
and their interference with the factorizable contributions should not be significant.
The comparisons between the measurement results with Belle's upper limit and phenomenological predictions are shown in Fig.~\ref{fig:BF_ratio_compare}.
Our results provide important input for these phenomenological models and contribute to the distinction between them.

The BESIII Collaboration thanks the staff of BEPCII, the IHEP computing center and the supercomputing center of the 
University of Science and Technology of China (USTC) for their strong support.
The authors are grateful to Yu-Kuo Hsiao for enlightening discussions.
 This work is supported in part by National Key R\&D Program of China under Contracts Nos. 2020YFA0406400, 2020YFA0406300; National Natural Science Foundation of China (NSFC) under Contracts Nos. 11335008, 11625523, 12035013, 11705192, 11950410506, 12061131003, 12105276, 12122509, 11635010, 11735014, 11835012, 11935015, 11935016, 11935018, 11961141012, 12022510, 12025502, 12035009, 12192260, 12192261, 12192262, 12192263, 12192264, 12192265, 12221005, 12225509, 12235017, 12005311; 
China Postdoctoral Science Foundation under Contracts No. 2019M662152, No. 2020T130636; the Fundamental Research Funds for the Central Universities, University of Science and Technology of China under Contract No. WK2030000053;
 the Chinese Academy of Sciences (CAS) Large-Scale Scientific Facility Program; 
 the CAS Center for Excellence in Particle Physics (CCEPP); 
 Joint Large-Scale Scientific Facility Funds of the NSFC and CAS under Contract No. U1732263, U1832103, U2032111, U1832207; CAS Key Research Program of Frontier Sciences under Contracts Nos. QYZDJ-SSW-SLH003, QYZDJ-SSW-SLH040; 100 Talents Program of CAS; The Institute of Nuclear and Particle Physics (INPAC) and Shanghai Key Laboratory for Particle Physics and Cosmology; ERC under Contract No. 758462; European Union's Horizon 2020 research and innovation programme under Marie Sklodowska-Curie grant agreement under Contract No. 894790; German Research Foundation DFG under Contracts Nos. 455635585, Collaborative Research Center CRC 1044, FOR5327, GRK 2149; Istituto Nazionale di Fisica Nucleare, Italy; Ministry of Development of Turkey under Contract No. DPT2006K-120470; National Research Foundation of Korea under Contract No. NRF-2022R1A2C1092335; National Science and Technology fund of Mongolia; National Science Research and Innovation Fund (NSRF) via the Program Management Unit for Human Resources \& Institutional Development, Research and Innovation of Thailand under Contract No. B16F640076; Polish National Science Centre under Contract No. 2019/35/O/ST2/02907; The Swedish Research Council; U. S. Department of Energy under Contract No. DE-FG02-05ER41374.

\bibliography{bibitem}

\end{document}

%% file: authorlist_2023-05-06.tex
\author{
\begin{small}
\begin{center}
M.~Ablikim$^{1}$, M.~N.~Achasov$^{5,b}$, P.~Adlarson$^{74}$, X.~C.~Ai$^{80}$, R.~Aliberti$^{35}$, A.~Amoroso$^{73A,73C}$, M.~R.~An$^{39}$, Q.~An$^{70,57}$, Y.~Bai$^{56}$, O.~Bakina$^{36}$
% ...
\\
\vspace{0.2cm}
(BESIII Collaboration)\\
\vspace{0.2cm} {\it
$^{1}$ Institute of High Energy Physics, Beijing 100049, People's Republic of China\\
% ...
$^{5}$ Budker Institute of Nuclear Physics SB RAS (BINP), Novosibirsk 630090, Russia\\
% ...
$^{35}$ Johannes Gutenberg University of Mainz, Johann-Joachim-Becher-Weg 45, D-55099 Mainz, Germany\\
$^{36}$ Joint Institute for Nuclear Research, 141980 Dubna, Moscow region, Russia\\
% ...
$^{39}$ Liaoning Normal University, Dalian 116029, People's Republic of China\\
% ...
$^{56}$ Southeast University, Nanjing 211100, People's Republic of China\\
$^{57}$ State Key Laboratory of Particle Detection and Electronics, Beijing 100049, Hefei 230026, People's Republic of China\\
% ...
$^{70}$ University of Science and Technology of China, Hefei 230026, People's Republic of China\\
% ...
$^{73}$ University of Turin and INFN, (A)University of Turin, I-10125, Turin, Italy; (B)University of Eastern Piedmont, I-15121, Alessandria, Italy; (C)INFN, I-10125, Turin, Italy\\
$^{74}$ Uppsala University, Box 516, SE-75120 Uppsala, Sweden\\
% ...
$^{80}$ Zhengzhou University, Zhengzhou 450001, People's Republic of China\\
\vspace{0.2cm}
% ...
$^{b}$ Also at the Novosibirsk State University, Novosibirsk, 630090, Russia\\
% ...
}
\end{center}
\vspace{0.4cm}
\vspace{0.4cm}
\end{small}
}